\documentclass[twocolumn,prb,floatfix,nobibnotes,nofootinbib]{revtex4-1}
\usepackage{hyperref}
\usepackage{amsmath}
\usepackage{bbold}
\usepackage{amssymb}
\usepackage{graphicx}
\usepackage{bm}
\usepackage{xcolor}
\newcommand{\editR}[1] {\textcolor{black}{#1}}
\newcommand{\editRR}[1] {\textcolor{black}{#1}}

\begin{document}

\title{A perspective on semiconductor-based superconducting qubits}

\author{R. Aguado}
\affiliation{\\
$^1$Instituto de Ciencia de Materiales de Madrid (ICMM), Consejo Superior de Investigaciones Cient\'{i}ficas (CSIC), Sor Juana In\'{e}s de la Cruz 3, 28049 Madrid, Spain. Research Platform on Quantum Technologies (CSIC).}

\date{\today}
\begin{abstract}
\editRR{Following the demonstration of semiconductor-based Josephson junctions which are fully tuneable by electrical means, new routes have been opened for the study of hybrid semiconductor-superconductor qubits. These include semiconductor-based  transmon qubits, single-spin Andreev qubits, and fault-tolerant topological qubits based on Majorana zero modes. In this perspective, we review recent progress in the path towards such novel qubit designs. After a short introduction and a brief digression about the historical roadmap that has led to the experimental state-of-the art, the emphasis is placed on superconducting qubits based on semiconductor nanowires.}
\end{abstract}\maketitle


\emph{Introduction}---During the last twenty years, we have witnessed an enormous progress, from fundamental physics to commercial applications, in what is known today as Quantum Technologies. This progress has brought a significant development in various areas including sensing, cryptography, telecommunications and computing \footnote{\url{https://ec.europa.eu/digital-single-market/en/policies/quantum-technologies-flagship}}. If we focus on the latter, there are various physical implementations that start to emerge as leading technologies. They include trapped ions, semiconducting devices and superconducting circuits. While it is is still too early to know which one will form the basis of tomorrow's quantum computers, superconducting qubits \cite{PhysRevA.69.062320,Devoret1169,Wendin_2017,Krantz,Kjaergaard} are arguably leading the race.  Research in this area is going beyond the academic sphere to enter the realm 
of big technology companies, such as Google and IBM, and make the news. However, despite this, going beyond the actual sizes, with 
a few tens of qubits forming part of so-called noisy intermediate-scale quantum processors (NISQ) to scalable quantum processors with thousands of qubits is far from trivial and represents an engineering challenge \cite{Kjaergaard}.  While so-called quantum supremacy can be demonstrated using NISQ processors for a dedicated task \cite{Nature-Supremacy}, full fault-tolerant quantum computing still requires a great deal of redundancy (many physical qubits for each logical qubit), to correct for decoherence.

\begin{figure*}
\centering \includegraphics[width=\textwidth]{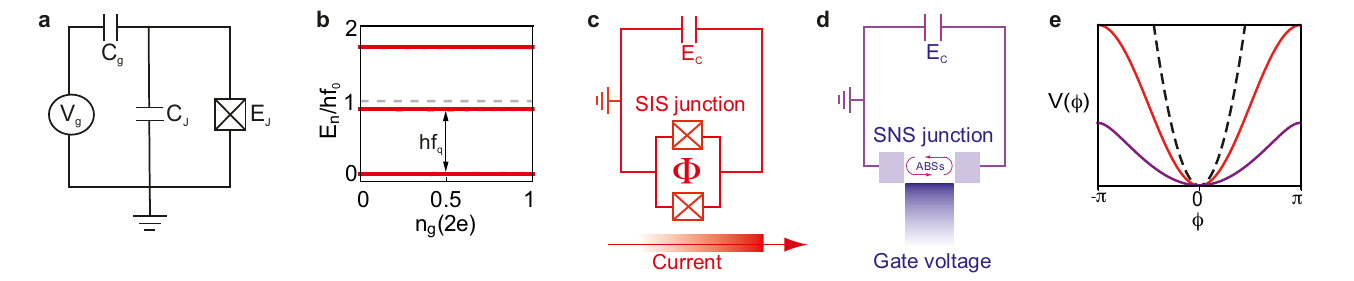}
\caption{\label{Fig1} a) Superconducting island circuit based on a Josephson junction (crosses) with Josephson coupling $E_J$ and capacitively shunted (charging energy $E_C=e^2/2(C_J+C_g)$).  b) In the transmon limit $E_J\gg E_C$ the spectrum is almost harmonic with a qubit frequency $f_q\approx f_{0}=\sqrt{8E_JE_C}/h$ and without charge dispersion $\partial E_n/\partial n_g\approx 0$, with $n_g=V_g/(2eC_g)$. c) Standard transmon based on a split Josephson junction (crosses) shunted by a capacitor (charging energy $E_C$). When both junctions are symmetric, the external flux $\Phi$ tunes the qubit frequency \editR{$f_{q}(\Phi)\approx\sqrt{8E_J E_C |cos (\pi\Phi/\Phi_0)|}/h$}. 
The external flux fixes the superconducting phase difference accross the junction $\phi=2\pi\Phi/\Phi_0$ where $\Phi_0=h/2e$ is the flux quantum . The flux is in turn modified by an external current line. d) Gate tuneable transmon (gatemon). The split SIS Josephson junction is replaced by a superconductor-normal-superconductor SNS junction, with the normal section being a gate-tuneable semiconductor material, containing Andreev bound states (ABSs). e) Phase-dependent potentials for the two junctions in c) and d): the standard SIS Josephson junction $V(\phi)=-cos\phi$ (red) and an arbitrary SNS junction containing Andreev bound states with $V(\phi)=-\sqrt {1-\tau sin^2 (\phi/2)}$ (purple, $\tau=0.95$ for the potential shown here). For reference, the dashed line shows a harmonic LC oscillator $V(\phi)=\phi^2$. \editR{Note that energy prefactors are not included in the potentials, which entails a trivial offset (different for each curve) to allow for better comparison around $\phi\approx 0$}}%
\end{figure*}

Among various superconducting qubit designs, the most relevant in the context of NISQ quantum computers is the transmon~\cite{koch2007charge}, a  qubit based on quantum degrees of freedom in circuits where Josephson junctions are shunted by large capacitors \editRR{(Fig. \ref{Fig1}a)} such that the Josephson energy $E_J$ of the junction exceeds the charging energy $E_C=e^2/2C$ of the shunting capacitor $C$. In this limit, the transmon spectrum is almost harmonic and results in a qubit frequency $f_{q}\equiv E_{10}/h$ which almost equals the Josephson plasma frequency  $f_{0}=\sqrt{8E_JE_C}/h$ (Fig. \ref{Fig1}b). The limit $E_J/E_C\gg 1$ exponentially suppresses charge dispersion and results in greatly enhanced coherence times~\cite{koch2007charge}. \editRR{State-of-the-art transmons in high-quality cavities reach coherence times in the range $50-100\mu s$ \cite{Kjaergaard}}. The trade-off is reduced anharmonicity, defined as the difference between the two first energy differences in the system $\alpha=E_{21}-E_{10}=-E_C$, \editRR{which is only a few percent of the qubit frequency ($\alpha\sim 100$ MHz versus $f_q\sim$ GHz)}. A small $\alpha$ reduces the operation time (namely, speed of operations) due to leakage out of the  qubit subspace, \editRR{with typical two-qubit gate times in the range of hundreds of nanoseconds \cite{Kjaergaard}}. The frequency of a transmon qubit can be tuned by using the magnetic flux that threads a loop geometry (Fig. \ref{Fig1}c). \editRR{This flux tuneability, which allows for faster gate operations, has however important drawbacks, the most important one being unwanted sensitivity to random flux fluctuations, known as flux noise \footnote{These fluctuations mostly appear due to fluctuations on the on-chip current bias that is applied to induce the flux and are typically  in the range $\approx (1\mu\Phi_0)^2/Hz$, with $\Phi_0=h/(2e)\approx 2\times 10^{-15}$Wb being the superconducting magnetic flux quantum (see Ref. \onlinecite{Krantz} for a detailed discussion about the different noise sources and their spectra in superconducting qubits)}}\editRR{, which reduces the typical relaxation and phase coherence times to the few $\mu s$ regime. 
Furthermore, the milliampere currents that control the flux generate crosstalk between qubits and heating, which limits integration \cite{3D}.} 

An interesting alternative is to try a change of paradigm by engineering hybrid combinations of materials. For instance, one could benefit from combining superconductors with semiconductors. The latter can be controlled by voltages that are less susceptible to heating and crosstalk. There are two main approaches when dealing with such hybrid semiconductor–superconductor circuits for quantum technologies. The first uses superconducting cavities (so-called circuit QED) to mediate quantum-coherent coupling over long distances of semiconductor qubits with long coherence times (usually electron spins in semiconductor quantum dots). While extremely powerful, this hybrid system does not represent a change of qubit paradigm per se, since the
superconducting cavity is just a mediator between standard spin qubits (a recent perspective about this approach has recently been published in Ref. [\onlinecite{doi:10.1063/5.0004777}]). The second approach, which is the subject of this article, also uses circuit QED but rather focuses on creating a novel type of hybrid qubit with potential advantages over standard superconducting qubits owing to its tuneability by electrical means (voltages). 

Specifically, the idea is to create qubits based on the quasiparticle degrees of freedom of hybrid semiconductor–superconductor Josephson junctions (JJ)s. Using this general concept, various designs have been proposed, e.g., Andreev qubits \cite{ZazunovPRL03,ChtchelkatchevPRL03}; parity-protected qubits, based on unusual $\pi$-periodic JJs \cite{Dou_ot_2012,larsen2020parityprotected}; or topological transmons \cite{Hassler_2011,Hyart:PRB13,ginossar2014microwave,PhysRevB.92.075143,10.21468/SciPostPhys.7.4.050,avila2020superconducting,avila2020majorana}, which combine transmon qubits with qubits based on Majorana zero modes (MZMs) \cite{Leijnse:SSAT12,Alicea:RPP12,Beenakker:ARCMP13,Sato:JPSJ16,Aguado:RNC17,Sato:ROPIP17,Lutchyn:NRM18,prada2019andreev,Aguado-Kouwenhoven:PT20,frolov:2019quest}. The latter are fault tolerant by virtue of their topological protection against local sources of noise \cite{Kitaev:PU01,Kitaev:AOP03,Nayak:RMP08}.

One of the key ingredients to realise such hybrid qubits is to have electrical control over the Josephson supercurrent that flows through the device. This can be accomplished by replacing the superconductor-insulator-superconductor (SIS) junction used in standard superconducting qubits by some other material with semiconducting properties, therefore allowing electrostatic gating (Fig. \ref{Fig1}d). Gate tuneable superconducting qubits, so-called gatemons, using this principle have recently been demonstrated with semiconductor nanowires (NWs) \cite{PhysRevLett.115.127001,Lange:PRL15}. Building on this progress \cite{gatemon-serious}, researchers are now using different device geometries and materials combinations to fabricate gatemon qubits compatible with high magnetic fields \cite{LuthiPRL18,Kroll2018,CharliePAT,sabonis2020destructive}, one of the essential prerequisites for reaching the topological regime. In what follows, we shall review the main concepts behind such hybrid qubits and the experimental state-of-the art.

\emph{Andreev levels}---
A unifying microscopic description of the Josephson effect in generic weak links considers how the superconducting pairing affects the non-superconducting region due to the proximity effect. The quantum coherent coupling between the two superconductor in a superconductor-normal-superconductor (SNS) junction depends on specific properties of the normal material in between.  This in turn is governed by Andreev reflection, whereby electrons are coherently retro-reflected back as holes at the NS interface while transferring a Cooper pair into the superconductor. Constructive interference between Andreev processes at both NS interfaces leads to coherent electron-hole superpositions, known as  Andreev bound states (ABSs) \cite{Kulik:SPJ70,FURUSAKI1991299}. Their energies depend on the superconducting phase difference $\phi$, but also on specific properties of the normal region, such as the transmission probability of electron transport across the weak link, and on the ratio between the length of the normal segment $L_N$ and the superconducting coherence length $\xi$. For a ballistic system, the latter reads $\xi=\hbar v_F/\Delta$, written in terms of the Fermi velocity $v_F$ of quasiparticles within the weak link and the superconducting gap $\Delta$. 

\editR{The simplest limit is the so-called short junction limit ($L_N\ll \xi$) at zero magnetic field and in the Andreev approximation (valid for high density systems, when the Fermi energy is much greater that $\Delta$). In this case, two (spin-degenerate) ABSs appear as subgap excitations in the junction at energies below the gap:
\begin{equation}
\label{ABS}
E_\mp(\phi)=\mp\Delta\sqrt{1-\tau \sin^2(\phi/2)},
\end{equation}
where $\tau$ is the normal transmission probability of a single channel (weak link) connecting the two superconducting electrodes. The electrodynamics of such Josephson weak link in a circuit depends not only on the ABSs energy but also on their occupation. Specifically, and since each of these ABSs carry opposite supercurrents \cite{RevModPhys.76.411}
\begin{equation}
\label{currrentABS}
I_\mp(\phi)=-\frac{2e}{\hbar}\frac{dE_\mp(\phi)}{d\phi}=\pm\frac{2e}{\hbar}\frac{\Delta}{4}\frac{\tau\sin(\phi)}{\sqrt{1-\tau \sin^2(\phi/2)}},
\end{equation}
the total current-phase relationship (CPR) reveals the occupation of ABSs. For instance, the ground state Josephson potential only contains the ABS below the gap $V_J(\phi)=E_-(\phi)$, which gives
a CPR
\begin{equation}
\label{currrentABS2}
I_J(\phi)=\frac{2e}{\hbar}\frac{dV_J(\phi)}{d\phi}=\frac{2e}{\hbar}\frac{\Delta}{4}\frac{\tau\sin(\phi)}{\sqrt{1-\tau \sin^2(\phi/2)}}.
\end{equation}
If, on the other hand, the junction contains one quasiparticle excitation, the Josephson potential is  $V_J(\phi)=E_-(\phi)+E_+(\phi)$, which gives $I_J(\phi)=0$. If a second quasiparticle populates the excited level, the Josephson potential is now $V_J(\phi)=E_+(\phi)$, which flips the sign of the CPR \cite{PhysRevLett.106.257003}.}

\editR{This simple example illustrates how a generic SNS junction, with many channels of arbitrary transparency \cite{Cayao:Belstein18} (and not necessarily at equilibrium), may strongly deviate from a standard SIS junction. The  SIS expression \cite{JOSEPHSON1962251,RevModPhys.46.251}
\begin{equation}
\label{SIScurrent}
 I_J(\phi)=I_c sin (\phi)=\frac{2e}{\hbar}E_J sin (\phi)
\end{equation}
is recovered in the (tunnel) limit where the junction contains many poorly transmitting channels, in which case the critical current follows the Ambegaokar-Baratoff formula \cite{PhysRevLett.10.486}, $I_c=\frac{\pi\Delta}{2e}G_N$, with $G_N$ being the normal conductance of the junction. Equivalently, this tunnel limit is described by a Josephson potential
\begin{equation}
\label{SISJosephson}
V_J(\phi)=-E_Jcos(\phi),
\end{equation}
which, in general, differs from the ABS expressions above (for a comparison, see Fig. \ref{Fig1}e).}

Importantly, the electron density in a semiconductor is tunable by external gates which enables electrostatic \editR{tuning of the junction, hence its ABSs,} and therefore of $V_J(\phi)$ and $I_J(\phi)$, thus creating gate-tunable Josephson elements. The gate-dependence of the number of conduction channels and of the set of transmission coefficients has been quantitatively characterised with great precision in NW JJs \cite{Goffman_2017}, \editR{using the multichannel generalization of Eq. \eqref{ABS}}. The CPR of NW JJs has been experimentally studied in  Ref. \onlinecite{SpantonNP:17}. Skewed CPRs, which strongly deviate from the $sin(\phi)$ in \eqref{SIScurrent}, indicate few-mode junctions with high transmissions which are in agreement with Eq. \eqref{currrentABS2}. Deviations are expected at low densities for junctions near pinch-off (where the Andreev-limit is no longer valid), at finite magnetic fields,  and in long junctions ($L_N/\xi\gtrsim 1$) \editR{containing many highly transmitting channels}. 

Direct visualization of the phase-dependent ABSs is also possible, although more challenging. This requires either tunnel spectroscopy using weakly coupled tunnel probes or microwave spectroscopy measurements. Some materials where direct ABS spectroscopy has been demonstrated include atomic break junctions \cite{BretheauNature13,Janvier:S15}, carbon nanotubes\cite{PilletNP10}, graphene\cite{BretheauNP17} and semiconducting NWs\cite{PhysRevLett.110.217005,vanWoerkomNP:17,HaysPRL:18,TosiPRX:19}. In NWs, the spin degeneracy of ABSs can be removed even at zero magnetic field \cite{TosiPRX:19} owing to the sizable spin-orbit interaction that couples occupied transverse subbands \cite{PhysRevB.96.125416}. The coupling of these spin degrees of freedom with the embedding superconducting circuit provides new avenues for quantum information processing. For example, coupling circuit QED devices to spin-polarised ABSs allows to use them as Andreev qubits \cite{ZazunovPRL03,ChtchelkatchevPRL03}. Similarly, this interplay can also be investigated in quantum dots embedded in superconducting circuits, where Yu-Shiba-Rusinov states emerge \cite{Lee:PRL12,Chang:PRL13,Lee:NN14,nyagard2016,Lee:PRB17,Su:NC17,Grove-Rasmussen:NC18,Su:PRL18}.

\begin{table*}
\begin{center}
\caption{Some milestones toward semiconductor-based superconducting qubits}
  \begin{tabular}{ | l | c | }
    \hline
Year	 &  $Milestone$   \\ \hline
1975   & First demonstration of Josephson effect in a semiconductor-based junction (thinned silicon wafers), Ref. \onlinecite{doi:10.1063/1.1655388}  \\ \hline
1979 & Superconductor contacted to p-InAs,  Ref. \onlinecite{1060276}   \\ \hline
1985 & 2DEG (inversion layer on a p-type InAs substrate) with Nb electrodes, some electrostatic gating of $I_c$ is observed, Ref. \onlinecite{PhysRevLett.54.2449}  \\ \hline
 1986 & SNS junction based on a n-InAs/GaAs heterostructure contacted to Nb electrodes, Ref. \onlinecite{doi:10.1063/1.97233} \\ \hline
 1995  & Supercurrent quantization in a split-gate InAlAs/InGaAs semiconductor heterostructure with Nb electrodes.  Ref. \onlinecite{PhysRevLett.75.3533} \\ \hline
 1997  & Evidence for a proximity-induced energy gap and multiple Andreev reflections in Nb/InAs/Nb junctions, Ref. \onlinecite{PhysRevB.55.8457}  \\ \hline
  2005& First demonstration of gate tunability of $I_c$ in JJs based on semiconductor NWs (Al/InAs) , Ref. \onlinecite{Doh272}  \\ \hline
  2015& Epitaxial growth of a thin aluminium shell on InAs, Refs. \onlinecite{KrogstrupNM:15,ChangNN:15,PhysRevB.93.155402}. \\ \hline
  2015& First demonstration of NW-based gate tunable transmon qubits (gatemons), Refs. \onlinecite{PhysRevLett.115.127001,Lange:PRL15}. \\ \hline
  2020& Epitaxial growth of various combinations of superconductors and semiconductors is demonstrated, Refs. \onlinecite{doi:10.1002/adma.201908411,khan2020transparent}. \\ \hline
 \end{tabular}
 \label{Table}
\end{center}
\end{table*}

\emph{Short historical background}---Experiments trying to demonstrate electrostatic tuning of the Josephson effect are by no means new. While Josephson's prediction was confirmed in less than one year in metal oxide tunnel junctions \cite{PhysRevLett.10.230}, it took almost a decade until the first convincing experimental demonstration of supercurrents in semiconducting junctions (thinned silicon wafers) was published \cite{doi:10.1063/1.1655388}. Historically, other important milestones (see Table \ref{Table}) include the first successful fabrication of superconducting contacts  deposited on p-type InAs in the late 1970s \cite{1060276} and the first demonstration in the mid 1980s of proximity effect in a two-dimensional electron gas (2DEG) formed in the inversion layer of a p-type InAs substrate coupled to Nb electrodes \cite{PhysRevLett.54.2449}. These experiments also showed some electrostatic gating of $I_c$ (the feasibility of the so-called Josephson junction field effect
transistor was discussed even earlier \cite{doi:10.1063/1.327935}). A similar experiment with a n‐InAs/GaAs heterostructure with Nb contacts was published soon after \cite{doi:10.1063/1.97233}. Further improvement was demonstrated ten years later, with the first experiments, realised in a split-gate InAlAs/InGaAs semiconductor heterostructure, demonstrating \cite{PhysRevLett.75.3533} the theoretical prediction \cite{PhysRevLett.66.3056,PhysRevLett.67.132} of supercurrent quantization in superconducting quantum point contacts. Evidence for a proximity-induced energy gap and multiple Andreev reflections in Nb/InAs/Nb junctions was reported soon after \cite{PhysRevB.55.8457}. This fast progress however encountered a serious bottleneck in the improvement of interface transparency, as reviewed in Ref. \onlinecite{Schapers:01}. 
This led to a valley of death with little improvement in the 2DEG-superconductor platform. For instance, experiments with ballistic InAs 2DEGs demonstrated very low critical currents related to an overall diffusive superconducting transport \cite{PhysRevB.60.13135}. Despite these difficulties, many strategies were explored to achieve high junction transparency. These, except a few promising results, such as e.g. the demonstration that silicon-engineering of Al/InGaAs heterostructures removed the native Schottky barrier and significantly increased the transparency \cite{doi:10.1063/1.122926}, proved unsuccessful.

A few years later, the field started afresh when the focus moved to semiconductor NWs coupled to superconductors. The first measurements demonstrating gate tunability of $I_c$ in JJs based on semiconductor NWs (Al/InAs) were published in 2005 \cite{Doh272}. Since then, other experiments have been reported  with various materials combinations including Al/SiGe \cite{XiangNN:06}, Al/InSb \cite{NilssonNL:12} and Nb/InAs \cite{Gharavi_2017}. The standard figure of merit for JJs is the product
$I_cR_N$, with $R_N=G_N^{-1}$ the normal resistance of the junction. This product is related to the gap by the relation $I_cR_N=\eta\Delta/e$, where the prefactor varies from $\eta=\pi$ (ballistic limit) to $\eta=\pi/2$ (diffusive limit) for transparent junctions.\cite{RevModPhys.76.411,PhysRevLett.10.486,RevModPhys.51.101}. Typical values in these first experiments were generally much lower than these upper bounds, still indicating imperfect proximity effect (poor Andreev reflection at the interface).  

An important breakthrough that greatly improved the interface transparency was the fabrication of heterostructures where a thin aluminium shell is epitaxially  grown on InAs, both in NWs \cite{KrogstrupNM:15,ChangNN:15} and in 2DEGs \cite{PhysRevB.93.155402}. Very recently, etching-free shadow epitaxy has successfully been implemented with other materials combinations beyond Al/InAs such as vanadium \cite{Bjergfelt_2019} and lead \cite{kanne2020epitaxial}; as well as other superconductors and semiconductors \cite{doi:10.1002/adma.201908411,khan2020transparent,pendharkar2019paritypreserving}. This new generation of devices shows $I_cR_N$ products close to the ideal limit, see e.g.  Ref. \onlinecite{khan2020transparent}, which demonstrates very good junction properties with very high transparencies. In parallel to this research (mostly focused on NW-based platforms), a great effort is being devoted to hybrid heterostructures  based on semiconducting 2DEGs and quantum well heterostructures \cite{PhysRevApplied.7.034029,Mayer19,Vigneau19,SrijitNC:19,PhysRevLett.124.226801}.

\emph{Hybrid Semiconducting-Superconducting Qubits}---Arguably, the most exciting possibility of hybrid semiconductor–superconductor devices is to create fault-tolerant topological qubits \cite{Kitaev:PU01,Nayak:RMP08,Sarma:NQI15} based on MZMs. These exotic excitations are predicted to emerge in a topological superconductor phase that can be engineered in hybrid semiconductor-superconductor materials by proper gate and magnetic field tuning \cite{Leijnse:SSAT12,Alicea:RPP12,Beenakker:ARCMP13,Sato:JPSJ16,Aguado:RNC17,Sato:ROPIP17,Lutchyn:NRM18,prada2019andreev,Aguado-Kouwenhoven:PT20,frolov:2019quest}. Moreover, the large $g$-factor and spin–orbit coupling of the typical semiconductors used in hybrids, InAs and InSb, are also crucial in order to reach the topological regime. Topological superconductivity can also be induced in planar hybrid JJs based on 2DEGS by tuning of the superconducting phase difference \cite{Fornieri:Nature19} and in full-shell geometries \cite{Vaitiekenaseaav3392,Penaranda:19}, where a superconducting shell fully coating the NW semiconducting core allows to induce full phase windings, akin to vortices.

\begin{figure}
\centering \includegraphics[width=\columnwidth]{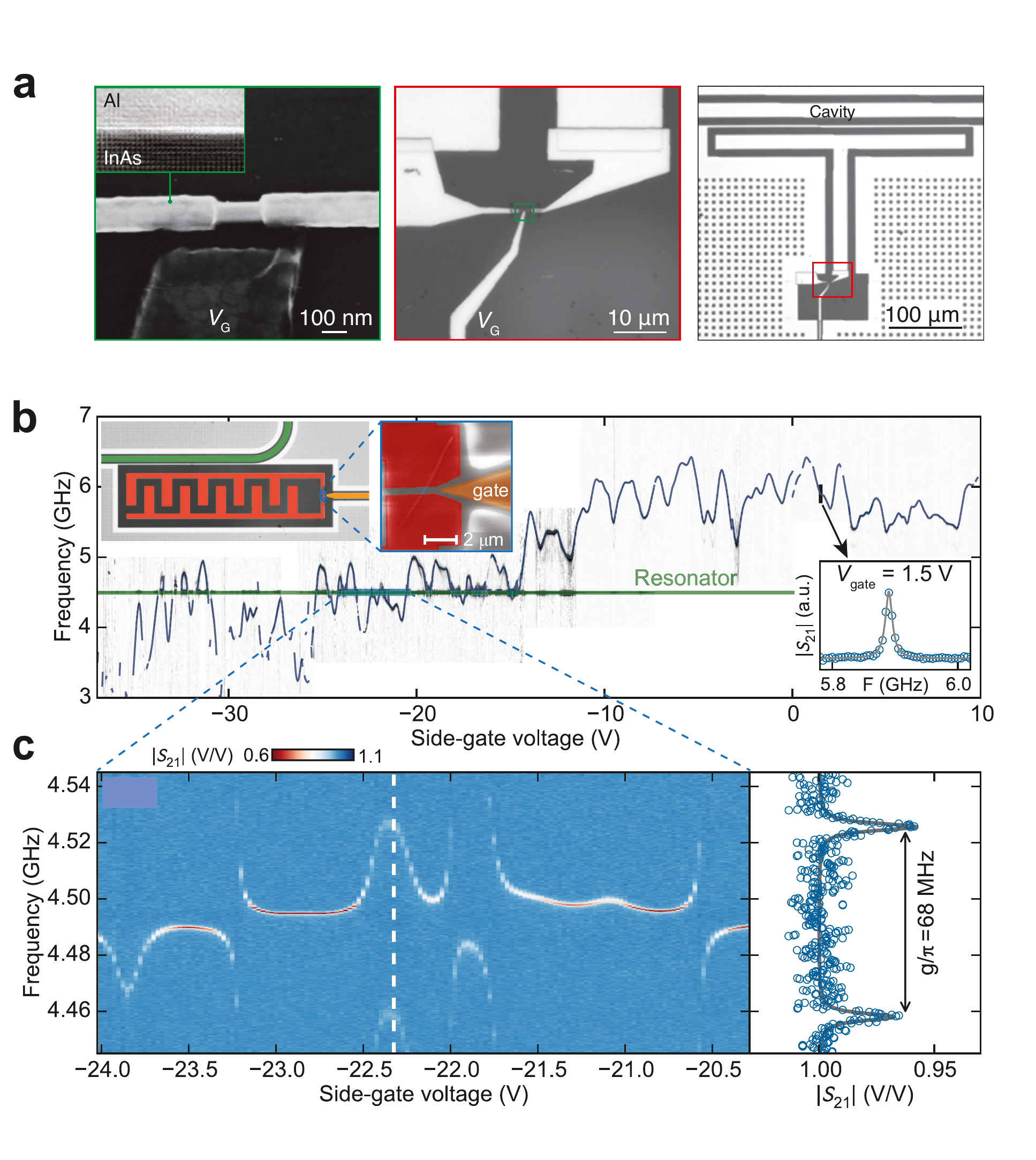}
\caption{\label{Fig2}   (a) Left: Scanning electron micrograph of a semiconductor-superconductor (InAs/Al) Josephson junction. The gate-tunable semiconducting weak link is created in the center, where a segment of the epitaxial Al shell is etched. This weak link can be depleted by means of a gate voltage $V_G$. The inset shows a transmission electron micrograph of the epitaxial InAs/Al interface. Center and right: the nanowire JJ is shunted by the capacitance of the T-shaped island to the surrounding ground plane. The whole structure is coupled to a transmission line cavity. Adapted from Ref. \onlinecite{PhysRevLett.115.127001}. (b) Top inset: image of a gatemon device showing the NW JJ and the proximal side gate (orange) for voltage control. Sweeping this side gate voltage induces reproducible changes in the qubit frequency. Lower inset: Example of the qubit transition at fixed voltage. (c) A blow-up around $V_g=-22 V$ shows multiple avoided crossings between the qubit and the resonator (green line in a). At $V_g=-22.3 V$ (white dashed line), the NW circuit fully hybridizes with the resonator. From the minimum splitting, a coupling strength between the NW circuit and the resonator of  $g/\pi\approx 68 MHz$ can be extracted. Adapted from Ref. \onlinecite{Lange:PRL15}. }
\end{figure}

The first steps towards these exotic qubits are already being demonstrated. Gatemon superconducting qubits using semiconducting JJs have been successfully implemented  with NWs \cite{PhysRevLett.115.127001,Lange:PRL15,PhysRevLett.116.150505,PhysRevB.97.060508,PhysRevLett.120.100502,PhysRevB.99.085434,PhysRevLett.124.056801}. An example is shown in Fig. \ref{Fig2}a, where the JJ is formed by an InAs NW partially covered by an epitaxial Al shell \cite{PhysRevLett.115.127001}. The supercurrent across the JJ can be tuned by the gate voltage ($V_G$ in left panel). This JJ is part of a larger circuit, including a shunting capacitor and the coupling to a transmission line (LC resonator), see central and right panel. Apart from the NW platform, similar architectures have successfully been demonstrated in 2DEGs \cite{Casparis2018,PhysRevLett.124.226801}, van der Waals heterostructures and graphene \cite{Kroll2018,Schmidt2018,Wang2019,Tahan19,Lee19}.  As we already advanced, an important prerequisite, apart from electrostatic tuneablity, is compatibility with high magnetic fields in the Tesla range. Examples include the niobium alloy NbTiN \cite{ZuoPRL17,AttilaNbTiN}, vanadium \cite{LeeNN14}, Ti/Al \cite{TiiraNC:17}, tantalum \cite{Estradaaeaav1235} and lead \cite{kanne2020epitaxial}. This has to be compared with standard SIS junctions, where the low critical field of the bulk superconductor (around 10 mT for aluminum) prevents reaching this regime. Another interesting strategy to greatly enhance the low critical field of aluminium \editR{is to reduce the aluminum layer thickness \cite{doi:10.1063/1.1659648}, something which is accomplished in the epitaxial shell heterostructures that we described before \cite{KrogstrupNM:15,ChangNN:15,PhysRevB.93.155402}.}

While gatemon qubits are clearly behind standard transmons, a great deal of progress have been made during the last few years. The first experiments \cite{PhysRevLett.115.127001,Lange:PRL15} have demonstrated electrostatic control of the Josephson energy
by a voltage gate that depletes carriers in the semiconducting weak link region. Changes in $E_J(V_g)=\hbar I_c(V_g)/2e$, result in shifts of the qubit frequency $f_{q}(V_g)\approx\sqrt{8E_J(V_g)E_C}/h$ over a wide range, from hundreds of Mhz to several Ghz, when the voltage of a nearby gate electrode changes in the few volt range. Fig. \ref{Fig2}b  illustrates this behaviour where the gatemon frequency shows reproducible gate-dependent fluctuations in the range $f_{q}(V_g)\approx 3-6GHz$ \cite{Lange:PRL15}. The inset shows the main qubit resonance at a fixed gate $V_g=1.5V$. Interestingly, the linewidth of the qubit transition $\gamma/\pi\approx 13MHz$ appears to be uncorrelated to the charge dispersion  $|\partial f_{q}(V_g)/\partial V_g|$, since a vanishing charge dispersion does not correlate with a reduction of $\gamma$. This seems to suggest that the main decoherence channel \editRR{is not yet dominated by charge fluctuations \footnote{Here, by charge fluctuations, we mean noise generated by charge traps of various origins (tunnel barrier, substrate dielectrics, etc). In standard superconducting qubits, it is usually modelled as a combination of $1/f$ noise at low frequencies and quantum (Nyquist) noise that progressively takes over at high frequencies in the GHz regime \cite{PhysRevLett.93.267007}. The authors of Ref. \onlinecite{Lange:PRL15} speculate that this lack of correlation between charge dispersion and linewidth could be attributed to quasiparticles populating the induced (soft) superconducting gap as opposed to better epitaxial samples with hard gaps like the ones in Ref. \onlinecite{PhysRevLett.115.127001}}}. 

Readout and manipulation of qubits can be accomplished by embedding the qubit in a superconducting cavity and using the toolbox of circuit QED, as in standard transmons. Strong coupling to the on-chip microwave cavity and coherent qubit control via gate voltage pulses has been accomplished in various platforms. The first effect is illustrated in Fig. \ref{Fig2}b, where strong coupling to the nearby LC resonator results in 
multiple avoided crossings when the gate voltage brings into resonance the qubit frequency $f_{q}(V_g)$ with the resonator frequency (green line). Fig. \ref{Fig2}c shows one of these avoided crossing of order $g/\pi\approx 68MHz\gg \gamma/\pi$. 
Other experimental achievements include qubit initialization, microwave-induced coherent quantum superpositions and measurements of relaxation and coherence times of around $\sim 1\mu s$, which greatly exceed gate operation times \cite{PhysRevLett.115.127001} . 
Subsequent work \cite{PhysRevLett.116.150505} has demonstrated various two-qubit quantum gates, including coherent swap operations and phase gates. A voltage-controlled quantum bus has also been demonstrated with gatemons \cite{PhysRevB.97.060508}. In these experiments, the superconducting cavity acts as a quantum bus, with a qubit-qubit coupling that can be controlled by an external gate voltage. Interestingly, the gate voltage can greatly enhance the coupling without reducing the qubit coherence properties. Another important development is the demonstration of controlled dc monitoring of gatemons \cite{PhysRevLett.124.056801}. The semiconducting characteristics of the JJ allows to compare in the same device important qubit characteristics, such as frequency and relaxation time, with the intrinsic properties of the junction such as $I_c$ and $R_N$. Further characterization of the junction properties includes the recent demonstration of simultaneous measurements of phase-dependent supercurrents and ABS spectrum in highly transmissive InAs JJs \cite{PhysRevLett.124.226801} with forward-skewed CPRs.  
\editRR{As we mentioned, the semiconducting JJ platform has potential advantages over standard transmons in terms of scalability. Furthermore, some of the parameters in the current generation of devices seem to be close (or even competitive in some cases) to their transmon counterparts. This includes single-qubit fidelities above 99\% and two-qubit controlled-phase gates with fidelities around 91\% \cite{PhysRevLett.116.150505}. Errors in the phase gate are attributed to relaxation and systematic phase errors that can be minimized using pulse shaping techniques \cite{PhysRevLett.112.240504}. Moreover, the dephasing times in the few $\mu s$ regime are comparable to conventional flux-controlled transmon qubits. Taken together, such figures suggest that gatemons already constitute a viable alternative towards scalable quantum computing.} 
 \editRR{Despite this progress, however, a word of caution is in order since gatemons still present important drawbacks, notably a wide variability in specific junction properties. This lack of control in junction variability, together with gate noise, could become a serious impediment if massive scalability is pursued.
The recent demonstration of controlled top-down fabrication of gatemons from wafer-scale 2D electron gases \cite{Casparis2018} seems a very promising route towards scaling up beyond a few qubits and improve reproducibility. However, the reported qubit coherence times of order $\sim 2\mu s$ are still limited by dielectric loss in the substrate \cite{oliver_welander_2013}.}

 \editRR{While some parameters can be provided for a comparison between gatemons and standard transmons, no experimental benchmark exists (at the time of this writing) that systematically compares important figures of merit in both platforms. Important questions, like e.g. whether charge noise coupled into the JJ via the gate/substrate is easier to mitigate than flux noise, remain open. Some efforts in this direction have been pursued by the authors of Refs. \cite{Lange:PRL15,LuthiPRL18} who have studied gatemons in a split-junction geometry that allows both flux and gate tuneability \footnote{In Ref. \onlinecite{Lange:PRL15}, the nonsinusoidal CPRs in the NW JJs induce strong deviations from transmon behaviour in the split geometry: while near $\Phi\sim\Phi_0/2$ the system behaves as a flux qubit, it exhibits transmonlike behavior near zero applied flux.}. As opposed to Ref. \onlinecite{Lange:PRL15}, where, as we mentioned, charge dispersion sweet spots seem to be uncorrelated to noise reduction, the experiments in Ref. \onlinecite{LuthiPRL18} clearly show evidence of coherence sweet spots, where the first order sensitivity $|\partial f_{q}(V_g)/\partial V_g|$ vanishes and the dephasing times correspondingly peak. The main dephasing noise is $1/f$ like with typical values $\approx (26\mu V/\sqrt{Hz})^2$. Flux sweet spots $|\partial f_{q}(V_g)/\partial \Phi|\approx 0$, on the other hand, allow to extract white noise and $1/f$ noise contributions with values $\approx (60 n\Phi_0/\sqrt{Hz})^2$ and $\approx (13 \mu\Phi_0/\sqrt{Hz})^2$, respectively. These experiments also study the evolution of the gatemon in a parallel magnetic field up to 70 mT, with an overall qubit frequency suppression following the expected BCS relation, together with relaxation and dephasing times showing features of unknown origin in their magnetic field dependence.}

\begin{figure*}
\centering \includegraphics[width=\textwidth]{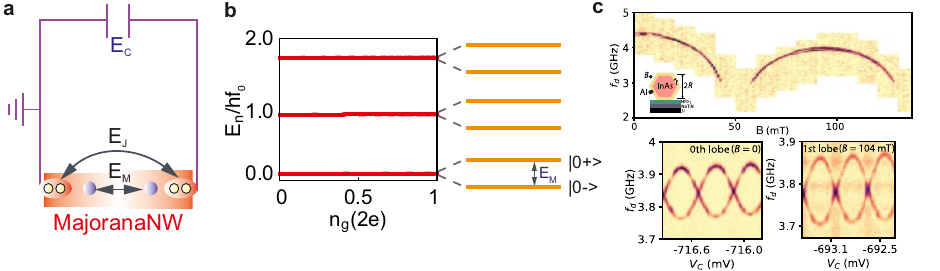}
\caption{\label{Fig3} \editR{(a) In a JJ containing Majoranas, apart from the standard Josephson effect (transfer of Cooper pairs with energy $E_J$), there is a single-electron coupling where a Cooper pair splits into two electrons and only one of these electron crosses the junction. As a result of this anomalous Josephson effect, each side of the JJ contains one extra electron (purple circles). This Majorana-mediated tunneling depends on an energy scale $E_M$. (b) As a result of Majoranas, each of the transmon lines splits, signalling coherent superpositions of island states with even/odd fermion parity. (c) Top: experimental two-dimensional map of two-tone spectroscopy as a function of drive frequency $f_d$ and parallel magnetic field $B$ for a gatemon based on a semiconducting nanowire fully wrapped by a superconducting shell (full-shell nanowire). The external flux threading the nanowire cross section (the inset shows a schematic with the hybrid InAs-Al nanowire and the substrate) modulates the superconducting gap (Little-Parks effect) which in turn modulates the qubit frequency. Around half-flux $\Phi\sim\Phi_0/2\sim 50mT$, superconductivity is fully destroyed; while at larger fields it is recovered with a slightly reduced qubit frequency. Bottom: gate dependence of the qubit frequency at zero and finite magnetic fields, showing even and odd fermionic parity branches. $B=104mT$ corresponds to a flux $\Phi\sim\Phi_0$ (first lobe), a regime where topological superconductivity has been predicted and reported in similar NW devices \cite{Vaitiekenaseaav3392}. The absence of clear avoided crossings between both fermionic parity branches, $E_M\sim 0$, could be due to an absence of zero modes, negligible coupling across the junction and/or absence of topological protection if zeros modes are present. Adapted from Ref. \onlinecite{sabonis2020destructive}}}%
\end{figure*}

\emph{Outlook and future directions: towards fault tolerant hybrid semiconductor-superconductor qubits}---In a JJ based on topological superconductors, hybridization of MZMs results in single electron transfers across the junction, as opposed to tunneling of Cooper pairs in a standard JJ. In this Majorana-mediated Josephson effect, a Cooper pair splits into two electrons, leading to the process $|0,0\rangle\equiv |n_L=0,\,n_R=0\rangle  \Longleftrightarrow |1,1\rangle\equiv |n_L=1,\,n_R=1\rangle $, where $n_{L/R}$ are the occupations at each side of the JJ (Fig. \ref{Fig3}a). This new Josephson term depends on an energy scale $E_M\sim\Delta_T\sqrt{\tau}$, with $\Delta_T$ being the topological gap that separates the Majorana modes from the rest of quasiparticle excitations in the junction and $\tau$ the normal transmission of the topological channel \cite{PhysRevB.79.161408,PhysRevLett.107.177002,Cayao:Belstein18,avila2020superconducting}. The minimal model that describes transmons based on such topological superconductor JJs takes the standard Josephson potential in Eq. \eqref{SISJosephson} and includes an extra term \
\begin{equation}\label{MajoranaJJ}
H_M=E_M\cos\phi/2.
\end{equation}
In few-channel junctions containing Majoranas, the ratio $E_M/E_J\approx\frac{\sqrt{\tau} \Delta_T}{\tau \Delta}$ is non-negligible with introduces new effects in the superconducting island. In particular, the new term generates coherent superpositions of the two fermion parity states $|0,0\rangle$ and $|1,1\rangle$. \editR{In the limit when the Majorana term does not couple different transmon bands, $hf_0\gg E_M$,} the new states are just the bonding/antibonding-like superpositions $|m,\pm\rangle \sim |m,1,1\rangle\pm  |m,0,0\rangle$, where $m$ denotes the transmon index \cite{ginossar2014microwave,PhysRevB.92.075143,10.21468/SciPostPhys.7.4.050}. These coherent superpositions appear as split transmon lines (Fig. \ref{Fig3}b), which form the basis of the so-called Majorana-Transmon qubit \cite{ginossar2014microwave,PhysRevB.92.075143,10.21468/SciPostPhys.7.4.050,avila2020superconducting,avila2020majorana}.  Theoretically, the topological protection of MZMs leads to greatly enhanced coherence times in such hybrid qubits \cite{ginossar2014microwave,PhysRevB.92.075143}. In practice, a microscopic description beyond the above minimal model of Majorana-mediated single-electron Josephson effect is needed \cite{10.21468/SciPostPhys.7.4.050,avila2020superconducting,avila2020majorana}. In such cases, the full quasiparticle spectrum in the JJ needs to be taken into account, leading to a low-energy phase-dependent spectrum which, in general, differs from Eqs. \eqref{ABS}, \eqref{SISJosephson} and \eqref{MajoranaJJ}. Such microscopic description allows to study the full evolution of the qubit, from a standard transmon to a Majorana-transmon hybrid, as magnetic field increases \cite{avila2020superconducting,avila2020majorana}. From the perspective of circuit QED, the split lines can be detected by microwave spectroscopy. Their dependence on different NW microscopic parameters allows to extract detailed information about the presence of Majorana modes in the junction  \cite{avila2020superconducting,avila2020majorana}.

Although, to date, no direct experimental signatures of Majoranas have been reported in gatemons, a few promising results are already demonstrating the potential of hybrid semiconductor-superconducting devices at high magnetic fields. This includes the first experiments in an InAs/Al double-island device containing all the ingredients of a Majorana-Transmon device \cite{CharliePAT}. These experiments indicate that, as magnetic field increases, \editR{a sizable hybridization of single-electron states belonging to two opposite-parity branches can be extracted from microwave spectroscopy (microwave-induced quasiparticle excitations). While this hybridization is consistent with a low-energy subgap state, no conclusive evidence on its physical character, wether it is an ABS near zero energy or a MZM, can be drawn from the experimental data. Interestingly, this single electron coupling (of order $E_{1e}/h\approx 10GHz$) is much larger than the Josephson coupling $E_J/h\approx 0.5GHz$. Even within the Majorana interpretation, where $E_{1e}=E_M$, the simple picture of split transmon lines in Fig. \ref{Fig3}b is no longer valid, since $E_M\gg E_J$ (a full analysis of these hybrid superconducting islands in different parameter regimes, including large $E_M$, can be found in Refs. \onlinecite{avila2020superconducting,avila2020majorana}).}

\editR{Recently, gatemons based on full-shell NWs have also been investigated \cite{sabonis2020destructive}. In these devices, an axial magnetic field applied to the small-diameter cylindrical superconducting shell gives rise to the Little-Parks effect which is characterised by a gap modulation where full destruction and subsequent reentrance of superconductivity driven by the external flux can be accomplished \cite{PhysRevB.101.060507}. This flux modulation in turn translates into Little-Parks modulation of the gatemon qubit frequency which can be tuned by more than $\approx 1GHz$ (Fig. \ref{Fig3}c, top panel). The qubit spectrum as a function of the gate voltage shows the simultaneous appearance of both fermion parity branches indicating fast quasiparticle poisoning (Fig. \ref{Fig3}c, bottom panel). The degeneracy between both parity branches is not removed even at finite magnetic fields, signalling an absence of coherent single-electron coupling across the junction $E_M\approx 0$ (or at least below the experimental resolution limit of $E_M/h<10MHz$, much smaller than the Josephson coupling $E_J/h\approx 4.7GHz$). The absence of coherent single-electron coupling, particularly near the first lobe where one flux quantum $\Phi_0$ is threading the full-shell NW, is somewhat unexpected in view of the recent claim of topological superconductivity and Majoranas in similar devices \cite{Vaitiekenaseaav3392}. One possible explanation could be just an absence of zero modes in the junction. If, on the other hand, we assume that the junction contains MZMs, a $E_M\approx  0$ could be due to either due to a poorly transmitting topological channel $\tau\rightarrow 0$ and/or a negligible topological gap $\Delta_T$. 
This latter scenario would be compatible with the recent observation that MZMs in full shell NWs do not enjoy generic topological protection \cite{Penaranda:19}. Further effects, like e.g. quantum dot formation and Yu-Shiba-Rusinov physics leading to non-topological zero modes in the junction \cite{valentini2020fluxtunable}, could influence the single-electron coupling $E_M$. All of these open questions call for further theoretical studies of the Josephson current and gatemon spectra in full shell NW JJs.}

Another interesting development investigates how to induce a strong suppression of charge dispersion in gatemons by properly tuning the resonant conditions of quantum dots in the junction \cite{PhysRevLett.124.246803,PhysRevLett.124.246802}. These results offer yet another avenue to careful engineering hybrid superconducting qubits with vanishing charge dispersion and large anharmonicity. When the suppression of charge dispersion is highest, the experiments demonstrate improved qubit coherence times \cite{PhysRevLett.124.246802}, which could be exploited for controllable fabrication of charge-insensitive qubits.

While in Majorana-based qubits it is the parity of the number of electrons that is relevant for protection; for so-called $\pi-$periodic JJs it is the parity of the number of Cooper pairs on an island what is relevant \cite{Dou_ot_2012}. This form of parity protection (not based on Majoranas) has recently been explored by using split junctions based on gatemons \cite{larsen2020parityprotected}. Specifically, a symmetric split junction formed by two nanowire JJs can realise a $cos(2\phi)$ Josephson element when frustrated by a half-quantum of applied flux and after proper gate-tuning. Such anomalous Josephson potential reflects coherent transport of in units of $4e$ charge, that is, pairs of Cooper pairs. This results in qubits having doubly degenerate ground states that differ by the parity of Cooper pairs on the island (compare with the Majorana-transmon design where the doubly degenerate states differ by one $e$ unit). The $cos(2\phi)$  regime near one half-quantum of applied flux results in greatly enhanced relaxation qubit times as compared to the standard transmon regime (for zero applied flux).

From a somewhat different perspective, more progress is also being reported about investigations of the dynamics of ABSs in NW JJs \cite{HaysPRL:18}. These experiments demonstrate coherent manipulation of ABSs using microwave techniques in a circuit QED geometry, as well as real time tracking of the bound-state fermion parity. The latter allows to extract typical timescales for fermion parity switches of the order of $T_{par}\approx 160\mu s$. Even without an underlying topology, hybrid JJs can be operated as qubits by using quasiparticles trapped in ABSs. Recently, the first demonstration of the single-shot readout of the spin of an individual quasiparticle in an ABS has been reported \cite{Hays:NP20}. Quantum non-demolition measurements on such Andreev level qubit can be performed by means of readout of the spin-dependent resonator frequency in a circuit QED setup, which allows to monitor the quasiparticle spin in real time. \editR{The key point is to inductively couple the junction to the superconducting microwave resonator} by making use of the spin-dependent supercurrents carried by spin-split Andreev levels \cite{PhysRevB.96.125416,TosiPRX:19,PhysRevLett.125.077701,metzger2020circuitqed}.
 
A detailed understanding of quasiparticle trapping in these hybrid devices is essential, since parity lifetimes will limit both superconducting spin qubits based on Andreev states and Majorana-based topological qubits. This progress, together with the great deal of activity around different aspects of gatemon physics, and all the recent investigations of different materials combinations, paves the way for further developments in the field of superconducting qubits based on gate-tunable semiconductor Josephson elements. The time is ripe for a major breakthrough which would demonstrate the great potential of hybrid semiconductor-superconductor qubits over standard superconducting qubits.

\acknowledgements
I thank Eduardo Lee for a critical reading of the manuscript. Research supported by the Spanish Ministry of Science and Innovation through Grant PGC2018-097018-B-I00. Support from the CSIC Research Platform on Quantum Technologies PTI-001 is also acknowledged. Data Availability Statement: Data sharing is not applicable to this article as no new data were created or analyzed in this study.

\bibliography{biblio}

\end{document}